# Homogeneous Cu-Fe super saturated solid solutions prepared by severe plastic deformation.

X. Quelennec[1], A. Menand[1], J.M. Le Breton[1], R. Pippan[2], X. Sauvage[1]*

*1- University of Rouen, CNRS UMR 6634, Groupe de Physique des Matériaux, Faculté des Sciences, BP12, 76801 Saint-Etienne du Rouvray, France*

*2- Erich Schmid Institute of Material Sciences, CD-Laboratory for Local Analysis of Deformation and Fracture, Austrian Academy of Sciences, Jahnstraße 12, A-8700 Leoben, Austria*

* Corresponding author: Xavier Sauvage (xavier.sauvage@univ-rouen.fr)

A Cu-Fe nanocomposite containing 50 nm thick iron filaments dispersed in a copper matrix was processed by torsion under high pressure at various strain rates and temperatures. The resulting nanostructures were characterized by transmission electron microscopy, atom probe tomography and Mössbauer spectrometry. It is shown that α-Fe filaments are dissolved during severe plastic deformation leading to the formation of a homogeneous supersaturated solid solution of about 12 at.% Fe in fcc Cu. The dissolution rate is proportional to the total plastic strain but is not very sensitive to the strain rate. Similar results were found for samples processed at liquid nitrogen temperature. APT data revealed asymmetric composition gradients resulting from the deformation induced intermixing. On the basis of these experimental data, the formation of the supersaturated solid solutions is discussed.

# 1. Introduction

The formation of metastable phases by mechanical alloying (MA) has been studied extensively during the past twenty years [1]. A large amount of experimental but also theoretical work has been done to clarify the physical mechanisms of the formation of non-equilibrium supersaturated solid solutions (ssss) in binary systems with a positive heat of mixing. The Cu-Fe system has been widely investigated by many authors for a full range of composition using numerous experimental techniques such as X-ray diffraction (XRD) [2-5], Mössbauer spectroscopy [5,6], Differential Scanning Calorimetry (DSC) [2-4], High Resolution Transmission Electron Microscopy (HRTEM) [5, 7, 8], Atom probe Tomography (APT) [9]. However, the physical mechanisms underlying the deformation induced intermixing are still under debate. This might be due to the following reasons:

(i) Mechanically alloyed products are nanostructured powders that are difficult to observe directly by TEM or APT. Thus, data about composition gradients or about any atomic scale mechanism are usually not available.

(ii) Mechanically alloyed products are inherently subject to contamination. Even if contamination by oxygen and oxidation problems can be avoided by careful milling in Ar atmosphere, there is always an unknown amount of material that is transferred from the steel balls and the vial to the powder.

(iii) The exact relationship between milling parameters and physical parameters like the temperature of the powder during deformation, the strain rate or the total accumulated strain is not known.

(iv) Basic deformation mechanisms leading to the high strain rate plastic deformation of powders under the impact of a steel ball are unknown.

(v) The density of crystallographic defects like dislocation or vacancy in MA powder is unknown while they should play a critical role in the intermixing mechanism.

vi) Supersaturated solid solutions prepared by ball milling are not always fully homogeneous even for very long milling time [7, 9, 10]. This is probably an indirect consequence of the non uniform total strain sustained by powders.

Anyway, various mechanisms have been proposed in the literature to account for the deformation induced Cu-Fe intermixing. Most of them rely on thermodynamic arguments: during milling, nanoscaled domains are formed leading to an interfacial energy sufficiently high to overcome the positive heat of mixing [2, 11]. However, such an approach does not provide any atomistic mechanism to account for the kinetics of the reaction. Some authors argue that pipe diffusion in the stress field of dislocations could play a role [4], but they do not provide any experimental evidence. The role of dislocations is also underlying the so-called "kinetic roughening" model of Bellon and Averback [12]. Atoms are supposed to be shifted across interfaces by the shear of atomic glide planes. The final state would be determined by the balance between these forced jumps and decomposition due to thermal diffusion. Recent APT measurements on the Cu-Ag system seem to validate this approach [13]. There are also some models incorporating both kinetic and thermodynamic effects where the ballistic events are introduced as excess entropy [3, 14] but microstructural features are not taken into account. With a pure kinetic approach, some other authors proposed a theory based on enhanced mobility resulting from deformation induced point defects [15], but unfortunately this assumption has not been yet tested by experiments.

Thus, for a better understanding of the mechanically induced intermixing as observed in mechanically alloyed powders, it is necessary to design some model experiments where deformation parameters are fully under control. As proposed in our earlier work [16, 17], severe plastic deformation (SPD) of bulk Cu-Fe composites by High

Pressure Torsion (HPT) may lead to some significant intermixing. The deformation mechanisms are very different comparing to mechanical alloying. During HPT the strain rate is fully under control, it is constant, and much smaller than during the impact of steel balls on powder grains. Since the strain rate is small, the temperature does not significantly increases and remains almost constant. Last but not least, HPT processing requires bulk samples that are easier to analyze or observe (TEM, APT) and there is no contamination.

The aim of the present study was first to optimize the process designed in [17] to obtain large volume fraction of homogeneous Cu-Fe ssss in a bulk sample. The second purpose was to investigate the influence of deformation parameters (strain level, strain rate and temperature) on the mechanically induced intermixing of Cu and Fe to get new insights into the physical mechanisms. In particular, tests were performed at low temperature where the mobility of vacancies is though to be negligible. Mössbauer spectroscopy was used to quantify the average amount of solid solution, while local analyzing techniques like TEM and APT were used to observe microstructures and to visualize diffusion gradients at the atomic scale. On the basis of these data, the driving force and the kinetics of the intermixing are discussed.

## 2. Experimental

The Cu-Fe composite processed by HPT was fabricated by accumulative cold drawing in a way very similar to the one used in our previous study [17]. The procedure was however modified in order to obtain a nanoscaled composite, i.e. a copper based material with iron filaments thinner than 100nm. Indeed, we have shown that the size of Fe grains is a critical parameter that directly controls the driving force of the intermixing reaction: the smaller they are the stronger the mixing [17]. To investigate the influence of deformation parameters a large volume fraction of ssss is needed, therefore it is necessary to process by SPD a composite with nanoscaled grains. Pure copper (99.9%) and pure iron (99.95%) were assembled as follows: an iron rod (2mm diameter) was inserted in a copper tube (inner diameter 4mm, outer diameter 6mm) that was inserted in an iron tube (inner diameter 7mm, outer diameter 10mm) that was inserted in a Cu tube (inner diameter 12mm, outer diameter 14mm). This assembly was cold drawn down to a diameter of 1mm, and then it was cut in pieces that were inserted again in a copper tube (inner diameter 12mm, outer diameter 14mm). This restacking procedure was applied five times. The resulting material is a copper based composite with nanoscaled iron filaments aligned along the wire axis (Fig.1). The average thickness of iron filaments is 50 nm, but there is a large distribution ranging from 25 to 100 nm. During the drawing process, intermediate annealing treatments were performed at 650°C to recover the ductility and to allow further reduction of the composite diameter. These annealing treatments were applied once the reduction of the cross section area of the wire was about 90%. In the fifth step, the annealing temperature was reduced down to 550°C to avoid coarsening and globularisation of nanoscaled iron filaments. The volume fraction of iron in the final nanocomposite is 12%, but as shown in the Fig. 1, the filaments are not homogeneously distributed within the copper matrix. This is due to the restacking in copper tubes during the processing by accumulative cold drawing as described above. During the last drawing step of the composite, the process was stopped at a diameter of 8mm. Thin discs (thickness 0.8mm) were cut with a diamond saw, sand blasted and then processed by HPT with a pressure of 6.2GPa up to 25 revolutions at various strain rates and temperatures. Due to the high pressure applied during the torsion straining, the thickness of the discs after processing was only about 0.5mm.

The microstructures were characterized by Scanning Electron Microscopy (SEM) using a LEO FE1530 (secondary electron detector) and by TEM using a JEOL 2000FX microscope operating at 200kV. X-ray Energy Dispersive Spectroscopy (EDS) was carried out using an Oxford Instrument detector. For SEM observations, samples were mounted, mirror polished and etched using 5% $HNO_3$ diluted in Ethanol. For TEM observations, samples were cut out at distance of 3±0.5mm from the disc center, mechanically grinded down to 100μm, dimpled down to 20μm and thinned down to electron transparency by ion milling using a GATAN PIPS 691 at 5keV. Atomic scale analyses were performed by APT using a CAMECA advanced delay line detector [18]. Specimens were prepared by standard electropolishing techniques [17] so that the tip was located at a distance of 3±0.5mm from the disc centre. Samples were field evaporated in UHV conditions with femtosecond laser pulses (energy of about 5 $10^{-7}$J at 2 kHz and specimen temperature of 20K) [19]. Analysed volumes were reconstructed with a mean atomic volume 11.8 $10^{-3}$ $nm^3$ and a field factor (kF) of 200 $V.nm^{-1}$. Transmission $^{57}$Fe Mössbauer spectrometry was performed at room temperature with a conventional spectrometer using a $^{57}$Co source in a rhodium matrix. Experimental spectra were fitted using hyperfine parameters defined for the Cu-Fe system by Campbell and co-authors in [20].

## 3. Results

To check that HPT could lead to the formation of a large volume fraction of Cu-Fe ssss, the Cu-Fe nanocomposite was first subjected to 25 turns at room temperature (293K) with a rotation speed of 0.4 rev/min. These severe plastic deformation conditions correspond to a total shear strain of about 540 and a strain rate of $0.2~s^{-1}$ where the sample was characterized (i.e. at a distance of 3mm from the HPT disc centre). The microstructures of the composite before and after HPT processing are compared in the Fig. 2. On the image showing the composite before deformation (Fig.2(a)), the filamentary structure does not appear because this is a cross sectional view and thus filaments are aligned along a direction parallel to the beam axis. The morphology of these filaments is very complex (see also Fig. 1(a)). As reported in the literature for other drawn composites [21, 22], this feature results from the <110>-fibre texture of the bcc α-Fe phase that induces a plane-strain deformation of this phase in the fcc Cu matrix. The selected area diffraction (SAED) pattern (Fig. 2(b)) shows Debye-Scherrer rings characteristic of polycrystalline structures with a very small crystallite size. In the initial state, both fcc Cu and bcc α-Fe phases are detected within the microstructure. After 25 turns by HPT, nanoscaled grains with an equi-axed structure clearly appear (Fig.2 (c)). The grain size is in a range of 10 to 50nm and only Debye-Scherrer rings corresponding to the fcc Cu phase are detected on the SAED pattern (Fig. 2(d)). However, using EDS measurements, the average amount of Fe measured in this area was 12±2 at.%. Thus, during the HPT deformation, the original iron filaments are dissolved and a Cu-Fe ssss is formed. APT analyses were carried out to image at the atomic scale the distribution of Fe atoms within this nanostructure. As shown on the three-dimensional reconstruction of the analyzed volume, the distribution of Fe atoms is fairly uniform (Fig. 3(a)). The average Fe concentration in this volume is 12.6±0.1at.%, which is consistent with EDS measurements. The distribution of Fe concentration measured in a $1.2 \times 1.2 \times 1.2 nm^3$ sampling volumes was plotted to statistically compare it with a random distribution (Fig. 3(b)). Although there is a small deviation for the highest and the lowest concentration, this distribution fits very well a random Bernouilli distribution, indicating that the Cu-Fe ssss is homogeneous.

The influence of HPT processing parameters on the volume fraction of Cu-Fe ssss was investigated by Mössbauer spectroscopy. This technique provides some information about the local environment of Fe atoms. If Fe atoms exhibit a bcc structure (α-Fe phase), absorption spectra are characterized by a magnetic sextet, but if Fe atoms are in solid solution in the fcc Cu phase, absorption spectra are characterized by a paramagnetic doublet [5, 20]. Spectra recorded for various levels of deformation up to 25 turns at room temperature (293K) and for a strain rate of $0.2s^{-1}$ are displayed in the Fig.4. The paramagnetic contribution obviously increases as the level of plastic deformation increases, indicating that this parameter has a strong influence on the formation of the Cu-Fe ssss. One should note that before HPT processing, there is already a small paramagnetic contribution. It is attributed to Fe atoms that have diffused in solid solution in the Cu matrix during annealing treatments performed during the accumulative cold drawing process. The volume fraction of Fe atoms in solid solution in the fcc Cu phase was directly computed from the fits of Mössbauer spectra, assuming that the area of each contribution is proportional to the number of absorbing Fe atoms. Figure 5, the proportion of Fe atoms in solid solution is plotted as a function of the number of HPT turns for two different processing temperatures (293 and 77K) and two different strain rates (0.2 and $0.02s^{-1}$). The dissolution of Fe filaments is very sensitive to the total plastic strain (number of HPT revolution) for any temperature or strain rate that were tested. Processing at low temperature (77K) slightly slows down the dissolution but at the end, there is only 15% less Fe atoms in solid solution than at 293K. It is also interesting to note that a change of the strain rate by one order of magnitude (from 0.2 down to $0.02s^{-1}$) does not significantly increase the dissolution rate.

The understanding of the physical mechanisms underlying the dissolution of iron filaments and the formation of the Cu-Fe ssss requires some investigation of the microstructure evolution for various levels of deformation. Some data collected in the sample processed by five revolutions by HPT are shown on the Fig. 6. On the bright field TEM micrograph two distinct regions are shown (Fig. 6(a)). In the upper part (labelled 1), there is a filamentary structure corresponding to a bunch of filaments located in the first Cu tube used for the preparation of the composite by accumulative cold drawing. A part of this Cu tube is imaged in the lower part of the image where grains are more equi-axed with a size in a range of 100 to 200 nm (labelled 2). One should note that under the shear strain, the thickness of the filaments was significantly reduced comparing to the initial material. SAED patterns taken in these filamentary regions always display both fcc Cu and bcc $\alpha$-Fe phases. Such a region was analyzed by APT, and the 3D reconstructed volume clearly shows an iron filament with a thickness of only few nanometres (Fig. 6 b)). However, if it assumed that the Cu-Fe composite is homogeneously deformed during the HPT process, the thickness of Fe filaments should be theoretically less than one nanometre after 5 revolutions (see [17] for details). This indicates that the deformation is more pronounced in iron depleted zones (i.e. former Cu tubes used for restacking as described in the previous section). As exhibited on the concentration profile computed across this iron filament (Fig. 6(c)) it does not contain any detectable amount of Cu, but a significant amount of Fe was detected in the Cu matrix. Due to the roughness of the interface, the gradient in a window of 1nm on each side of the interface cannot be rigorously interpreted; however on the copper rich side a long range composition gradient spreading over 5 nm is clearly exhibited.

After 14 revolutions by HPT, the original lamellar structure hardly appears on TEM bright field images (Fig. 7(a)). As shown on the 3D APT reconstructed volume (Fig. 7(c)), the lamellae are very thin (down to few nanometres) and exhibit a wavy morphology. The composition profile computed across Cu/Fe interfaces (Fig. 7(b)) shows that the Fe concentration gradient in the fcc Cu phase is larger than after 5 revolutions (Fig. (6)). The fcc Cu phase contains up to 30 at.% Fe in solid solution while iron filaments do not seem significantly affected by Cu (excepted the thinnest filament on the right of the profile). Thus, in agreement with Mössbauer data, APT measurements confirm that the higher the level of deformation the larger the amount of Fe in solid solution in the fcc Cu phase.

The microstructure of the composite processed by HPT at 77K (liquid nitrogen temperature), was also characterized by TEM and APT. The microstructures look very similar and no significant difference was observed on the micrographs. After 14 revolutions, there are still some regions with a lamellar structure containing bcc $\alpha$-Fe as shown in Fig. 8(a) (region labelled 1). This image is very interesting because a region with nanoscaled equi-axed grains is also exhibited (labelled 2). The SAED pattern from this area clearly indicates that it is fully fcc, e.g. there is no $\alpha$-Fe phase. However, EDS measurements revealed that it contains 13.4 ±2 at.% Fe and thus that it is a fcc Cu-Fe ssss. The transition between these two regions and the mechanisms of the formation of the nanoscaled equi-axed ssss will be discussed in the next section.

APT measurements were performed after 25 revolutions at 77K (Fig. 8(b)). Like in the composite deformed at room temperature, this 3D volume exhibits a homogeneous distribution of Fe atoms (12.2±0.1 at.% ). Like in the sample processed at room temperature (Fig. 3), the distribution of Fe atoms was compared to a random distribution (data not shown here). No significant difference was found, confirming the homogeneity of the ssss.

**4. Discussion**

Homogeneous super saturated solid solutions (ssss) of about 12at.% Fe in fcc Cu were achieved by SPD. This is far beyond the equilibrium solubility limit at room temperature and even larger than the solubility of Fe in Cu at 1350K (about 4at.% [23]). Such ssss were already obtained in the past in the same system using ball milling of powders, however in the present study bulk samples were produced and as demonstrated by APT data the ssss is fully homogeneous which is usually not the case in ball milled powders [9, 10, 24]. It is however worth noticing that the sextet typical of $\alpha$-Fe still appears even in samples processed up to 25 revolutions by HPT. This can be simply explained by the deformation gradient in HPT discs: the shear deformation is a linear function of the radius, thus the cumulated strain in the disc centre is quite low and not strong enough to promote the dissolution of the $\alpha$-Fe phase. A lead shield (diameter 4mm) was used to mask the centre of the disc, and the recorded Mössbauer spectrum (data not shown here) did not exhibit the sextet, confirming that the $\alpha$-Fe phase is located only in the low strained region.

These data raise two questions: what are the physical mechanisms of the deformation induced intermixing, and why are the present super saturated solid solutions so homogeneous? The experimental data show that the mixing is not strain rate sensitive (in the range of 0.02 to 0.2 $s^{-1}$), it is reduced by about 15% if the temperature is decreased from 293K to 77K and it starts with asymmetric composition gradients. Two different approaches could be considered to understand the evolution of the nanostructure under severe plastic deformation: the first one involving thermodynamic destabilization and the second involving pure mechanical mixing.

*4.1 Thermodynamic destabilization*

The microstructure of the Cu-Fe composite investigated in the present study is characterized by a very small grain size and thus a huge proportion of interfaces. As suggested for ball milled powders, the driving force might be the high interfacial energy of the nanoscaled Cu-Fe mixture combined with the high density of crystalline defects [2, 4, 11]. It is also important to note that it was recently proposed by Kozeschnik that such intermixing may significantly reduce the Cu-Fe interfacial energy [25]. It is however important to note that grain growth or coarsening would have a similar effect, but the continuous deformation of the two phase mixture prevents such mechanism and even continuously increases the Cu/Fe interfacial area and thus the driving force for mixing. As suggested by Jiang and co-authors [5, 6], in such a highly strained and nanoscaled system, the bcc $\alpha$-Fe phase may also transform into fcc. Such a transformation may be promoted by the small grain size [26], the shear stress [27] or to decrease the interfacial energy like in the early stage of precipitation of Fe in fcc Cu where Fe precipitates exhibit the meta-stable fcc crystallographic structure [28]. However, this fcc Fe phase was not detected on the Mössbauer spectra (paramagnetic contribution at room temperature characterized by a singlet with an isomer shift of 0.09mm $s^{-1}$ [29]) or by TEM, in the present investigations.

With this scenario, the mixing rate and the composition gradients are not controlled by the driving force (interfacial energy) only. Of course, the atomic mobility plays a crucial role and the following condition is required: thermal diffusion should be significant or an alternative media may exist.

*(i) Thermal diffusion*

During ball milling, the local temperature following impacts of steel balls might be high enough to promote diffusion, but this is obviously not the case during HPT experiments. Indeed, the temperature is directly linked to the strain rate and our data clearly show that this latter parameter has little effect on the dissolution rate. Thus, this feature clearly indicates that, as expected, the temperature does not significantly increase during the torsion under high pressure. Since the atomic mobility of Cu in bcc α-Fe and of Fe in fcc Cu is not significant at room temperature or below [30, 31], thermal diffusion cannot account for the enhanced atomic mobility and other mechanisms have to be considered.

*(ii) Alternative diffusion media*

Nanostructured materials processed by SPD are characterized by non-equilibrium grain boundaries acting as fast diffusion path [32], high dislocation densities and high vacancy concentration [33-36]. However, only these two later kinds of crystalline defects could promote bulk mixing leading to the formation of homogeneous ssss. The role of dislocations as possible diffusion pipe has already been proposed for the mixing of ball milled powders [4, 37]. However, in the present study, no segregation along linear defects was observed by APT, even for the lowest deformation rate. The reason might be that in such nanostructured alloys, dislocations are emitted but also annihilated along boundaries during plastic deformation. Concerning SPD induced vacancies, the formation of point defects by plastic deformation has been reported since the 60's and various models are proposed in the literature [38-40]. The role of high internal stresses [41, 42], high strain rates [43, 44] and SPD [33-36] have been investigated. In SPD processed materials the vacancy concentration is typically in a range of $10^{-5}$ to $10^{-4}$. However, to enhance the atomic mobility not only a high density of vacancy is necessary, but vacancies should also exhibit a significant mobility. During the HPT process, the material is deformed under a pressure of 6.2GPa, corresponding to an elastic deformation of about 5% for the fcc Cu phase and 3% for the bcc Fe phase. In such conditions, as demonstrated by Sato and co-authors [42], the migration energy of vacancies may be reduced by 30%. As suggested in our earlier studies, such conditions sufficiently enhance the diffusion rate of Fe in fcc Cu to account for bulk diffusion at room temperature [17, 45]. However, the present investigations clearly show that at cryogenic temperature (77K) large amounts of homogeneous ssss can be obtained while the mobility of vacancies is not significant. Therefore, the following "pure mechanical mixing" mechanism should be considered.

*4.2 Pure mechanical mixing*

Following the approach proposed by Bellon and Averback based on Monte Carlo simulation [12] and recently supported by atomic scale investigation of ball-milled Ag-Cu powders [13], one may consider the so-called "kinetic roughening" model. During the course of SPD, atoms may be shifted across interfaces by dislocation induced shear of atomic planes. Such forced atomic jumps may lead to a significant Cu-Fe intermixing near Cu/Fe interfaces. This mechanism is intrinsically symmetric, but asymmetric concentration profiles may result from the balance between these forced jumps and the decomposition due to thermal diffusion. Considering such a mechanism for the Cu-Fe system of the present study, it seems that Cu atoms mechanically forced to diffuse in the α-Fe phase are easily swept back to the fcc Cu phase while Fe atoms inserted in the fcc Cu phase do not have enough thermal mobility to fully balance strain induced atomic jumps. Then, at higher strain there is more and more Fe atoms in solid solution in the fcc Cu phase while α-Fe grains shrink and progressively disappear which gives rise to a homogeneous ssss.

With this scenario, the mixing mechanism and composition gradients are not only controlled by the strain level, but as well by the solubility limit (e.g. the driving force for decomposition) and the thermal diffusion coefficients of Cu in bcc α-Fe and of Fe in fcc Cu. So, the following conditions are required:

*(i) Cu/Fe interfaces must be sheared by dislocations.*

From our TEM observations of the early stage of deformation, Cu/Fe interfaces always appear continuous in all directions (cross sectional and longitudinal views). It seems that the continuous deformation by HPT, contrary to ball milling, does not promote significant shearing of Cu/Fe interfaces. As observed in cold drawn metal matrix composite, both phases co-deform along the flow direction [21, 22]. But for higher level of deformation when α-Fe filaments become extremely thin and break-up, the situation is probably very different. In principle, atomic scale analyses by APT should be able to exhibit sheared interfaces. The reason why it was never clearly observed could be due to the subsequent decomposition induced by thermal diffusion. This phenomenon may indeed strongly affect the topology of interfaces.

*(ii) Cu and Fe should exhibit different thermal diffusion coefficient or/and decomposition driving force.*

At low temperature, the Cu-Fe phase diagram is quite symmetric, and the solubility extremely limited on both the Cu rich and Fe rich side. At 1100K, the solubility limit is 4at.% Fe (respectively 2at.% Cu) in fcc Cu (respectively bcc Fe) [23]. So the driving force for decomposition might be higher in the bcc α-Fe phase. However, at 500K the thermal diffusion coefficient of Cu in bcc α-Fe is 5 orders of magnitude lower than that of Fe in fcc Cu [30, 31]. So, Cu ssss in bcc α-Fe should be more difficult to decompose than Fe ssss in fcc Cu, which is not consistent with APT data. However, as discussed in the following, the atomic diffusivity might be affected under HPT conditions.

*(iii) Thermal diffusion should be significant or an equivalent media may exist.*

As already discussed previously for the so-called "thermodynamic destabilization" mechanism, thermal diffusion cannot give rise to significant atomic mobility but SPD induced point defects may contribute to solute transport. It is interesting to note that positron lifetime spectroscopy experiments have demonstrated that deformation induced vacancies often agglomerate in bcc Fe while they are more homogeneously distributed in fcc Cu [46]. This could be attributed to a lower migration energy in the fcc Cu phase which in turn may promote the formation of the asymmetric composition gradients revealed by APT.

Considering these two mixing mechanisms and their related arguments, one may also imagine that a combined process occurs.

*4.3 Mixed mechanism*

Under a continuous and slow deformation process like HPT, dislocations could not simply cross Cu/Fe interphase boundaries and that is probably why we did not find any experimental evidence of this mixing mechanism (previously called mechanical mixing). However, if some Fe atoms diffuse in the fcc Cu phase to relax some interfacial energy (following the thermodynamic destabilization mechanism), concentration gradients would appear and consequently deeply modify the configuration of Cu/Fe interfaces. In such a situation, it seems reasonable to think that mechanical mixing would be more likely to occur. Dislocations stored along boundaries in the early stage of deformation and SPD induced vacancies may promote the formation of mixed interfaces. At cryogenic temperature, the lower atomic mobility would simply delay the process leading to a lower volume fraction of ssss at after 25 revolutions by HPT.

Last, one should note that ssss obtained by HPT are homogeneous with equi-axed nanoscaled grains (Figs. 2, 3 and 8). Thus, the dissolution of $\alpha$-Fe filaments is combined with a progressive destruction of the original filamentary structure of the Cu-Fe composite. The thickness of $\alpha$-Fe filaments in the course of deformation is only few nm (Figs. 6 and 7). This is much smaller than the mean grain size of the ssss (in a range of 10 to 50 nm, Fig. 2), therefore as already reported in pure copper [47], one may conclude that boundary migration (i.e. dynamic recrystallisation) occurred during the HPT process. In the early stage of deformation, the huge amount of Cu-Fe interfaces prevents any grain boundary motion [28], but once the $\alpha$-Fe phase is dissolved, dynamic recrystallisation and grain growth may freely occur to reduce the interfacial energy. One should note that similar features have been recently reported in other nanocrystalline materials processed by SPD [47-49] and it is though that the movement of boundaries may promote the chemical homogenisation process.

**5. Conclusions**

i) Fully homogeneous Fe ssss in fcc Cu were obtained by SPD under controlled deformation parameters (plastic strain, strain rate and temperature). Up to 12 at.% Fe atoms homogeneously distributed were dissolved in the fcc Cu phase. Once $\alpha$-Fe filaments are completely dissolved, some dynamic recrystallisation occurs and equi-axed nano-grains do form.

ii) The dissolution rate is mostly controlled by the accumulated plastic strain while it does not significantly change for a strain rate in range of 0.2 to $0.02s^{-1}$. SPD at liquid nitrogen temperature delay the dissolution but very similar ssss also forms.

iii) As shown by APT data, ssss results from Cu-Fe intermixing. During the early stage of deformation, Cu-Fe interfaces exhibit asymmetric composition profiles.

iv) Two different models were considered to account for the formation of ssss: "pure mechanical mixing" where the diffusion mechanism is controlled by shuffling of atoms induced by shear of atomic planes, and "thermodynamic destabilization" where the Cu/Fe interfacial energy is the driving force for $\alpha$-Fe dissolution in the fcc Cu matrix. It is concluded that the observed ssss may result from a combination of these two mechanisms.

# Figures

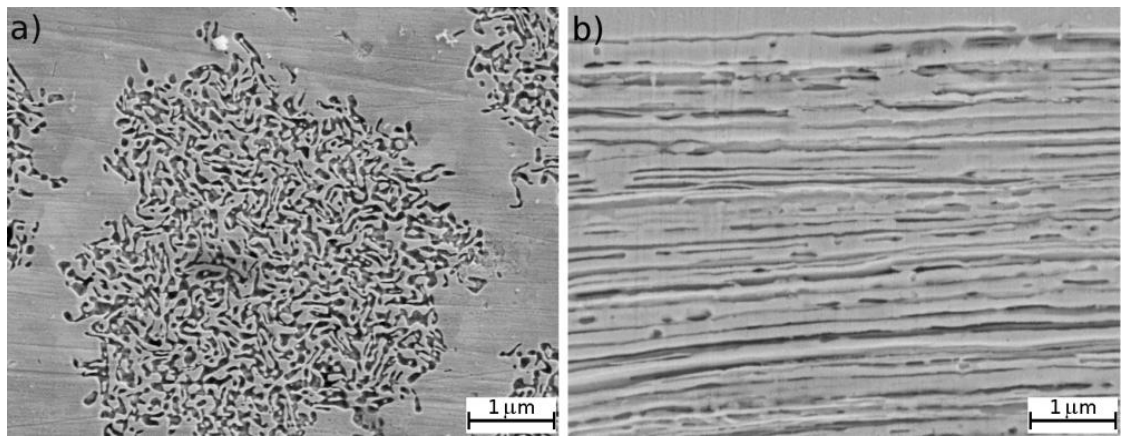

**Figure1.** SEM pictures of the Cu-Fe nanocomposite processed by accumulative cold drawing. Iron filaments are darkly imaged. (a) Cross sectional view of the wire showing a bunch of iron filaments stacked in the copper tube of the second step of the process (see text for details). The average thickness of iron filaments is 50 nm. (b) Longitudinal view of the wire showing the alignment of iron filaments along the wire axis.

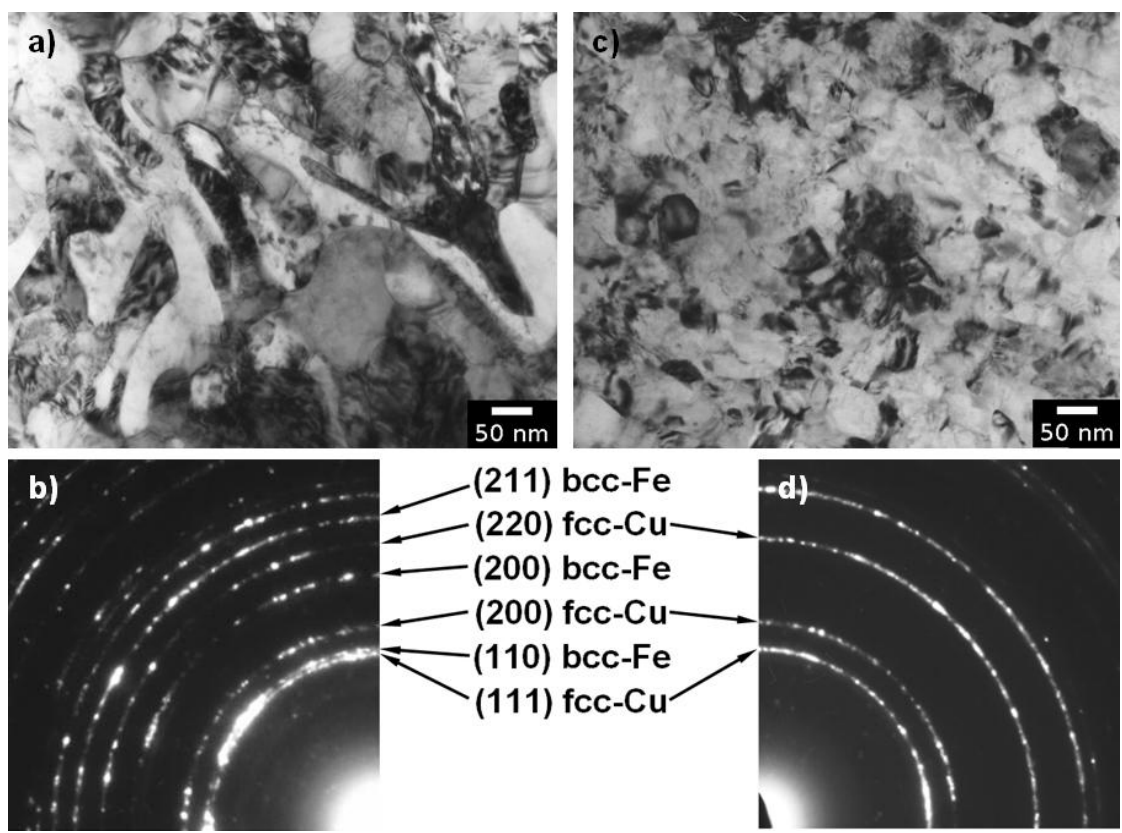

**Figure2.** (a) Bright field TEM image of the Cu-Fe composite before severe plastic deformation; (b) corresponding Selected Area Electron Diffraction (SAED) patterns where both fcc Cu and bcc α-Fe phases are detected. (c) Bright field TEM picture of the Cu-Fe composite after 25 revolutions by HPT at 293K with a strain rate of $0.2\ s^{-1}$ ; (d) corresponding SAED pattern where only the fcc Cu phase is detected.

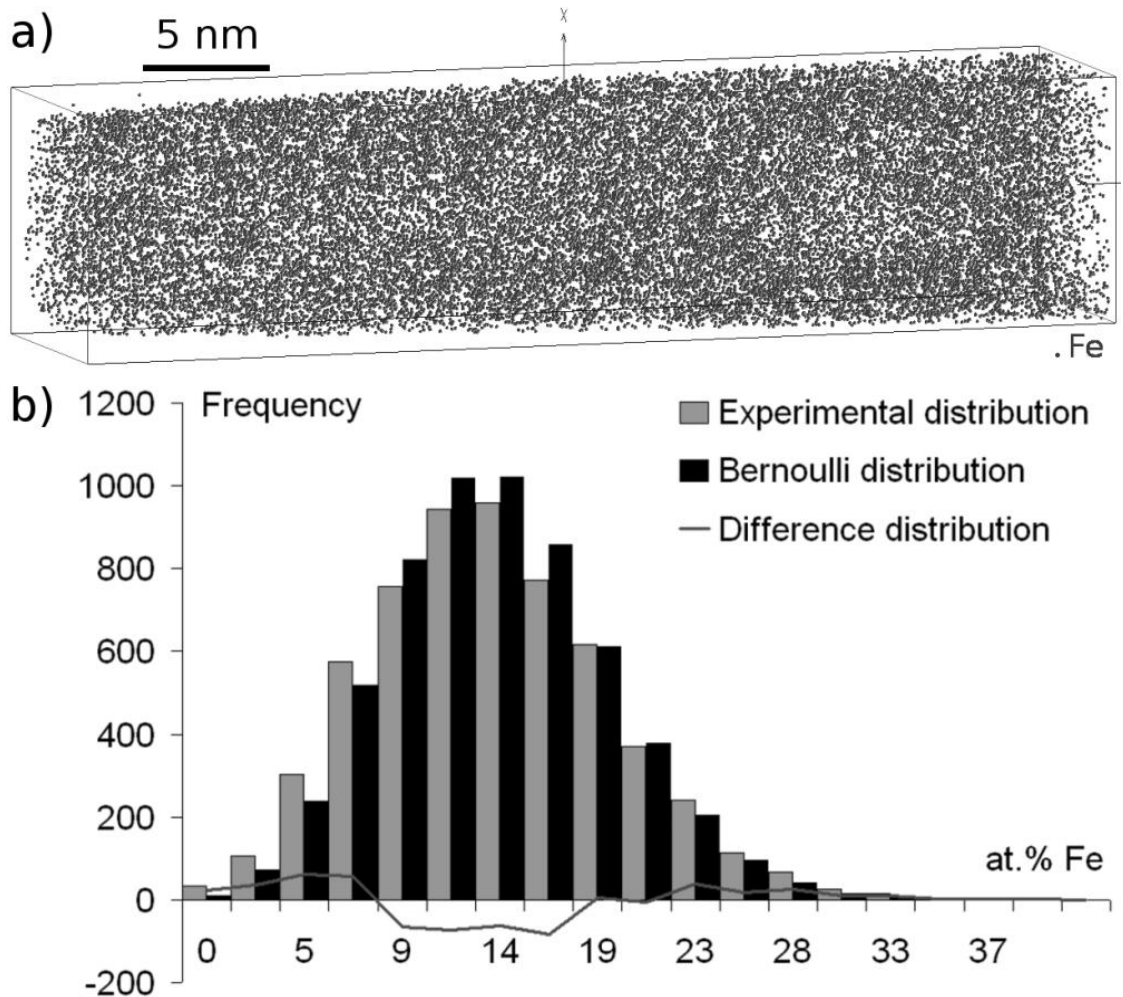

**Figure 3.** (a) Three-dimensional reconstruction of a volume analyzed by APT in the Cu-Fe composite processed by HPT (25 revolutions at 293K with a strain rate of 0.2 s$^{-1}$). Only Fe atoms are displayed to show their homogeneous distribution within the volume. The Fe concentration in this analysed volume is 12.6 at.% (±0.1). (b) Distribution of Fe concentrations computed in 1.2x1.2x1.2 nm$^3$ sampling volumes compared with a Bernoulli random distribution.

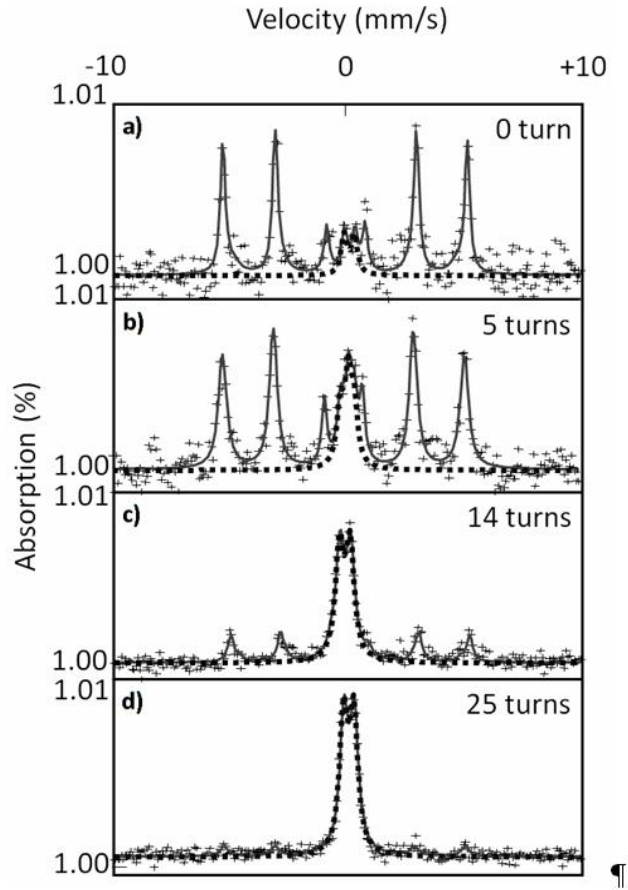

**Figure 4.** (a) Mössbauer spectra recorded on the Cu-Fe nanocomposite before HPT processing. (b) after HPT processing by 5 turns, (c) 14 turns and (d) 25 turns (at 293K and with a strain rate of $0.2s^{-1}$). The paramagnetic contribution corresponding to the Cu-Fe ssss is fitted in red while the black line is the sum of both the ferromagnetic (corresponding to the bcc $\alpha$-Fe) and the paramagnetic contributions used to fit the spectra.

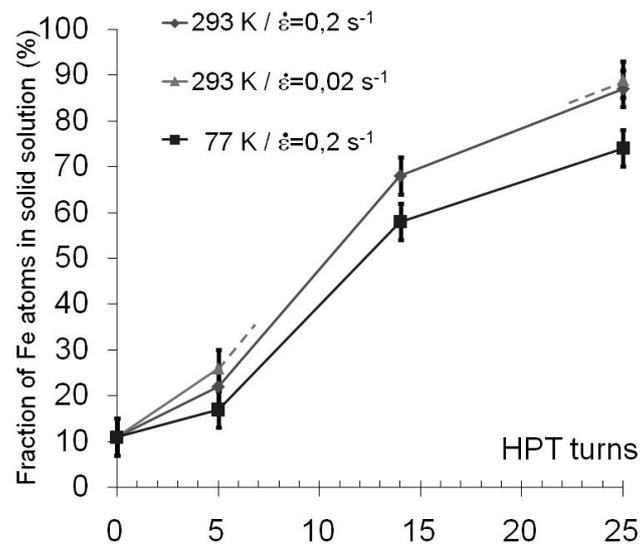

**Figure 5.** Volume fraction of Cu-Fe solid solution measured by Mössbauer spectroscopy plotted as a function of the level of deformation by HPT. Measurements were carried out at two different temperatures (77 and 293K), and two different strain rates (0.2 and $0.02s^{-1}$).

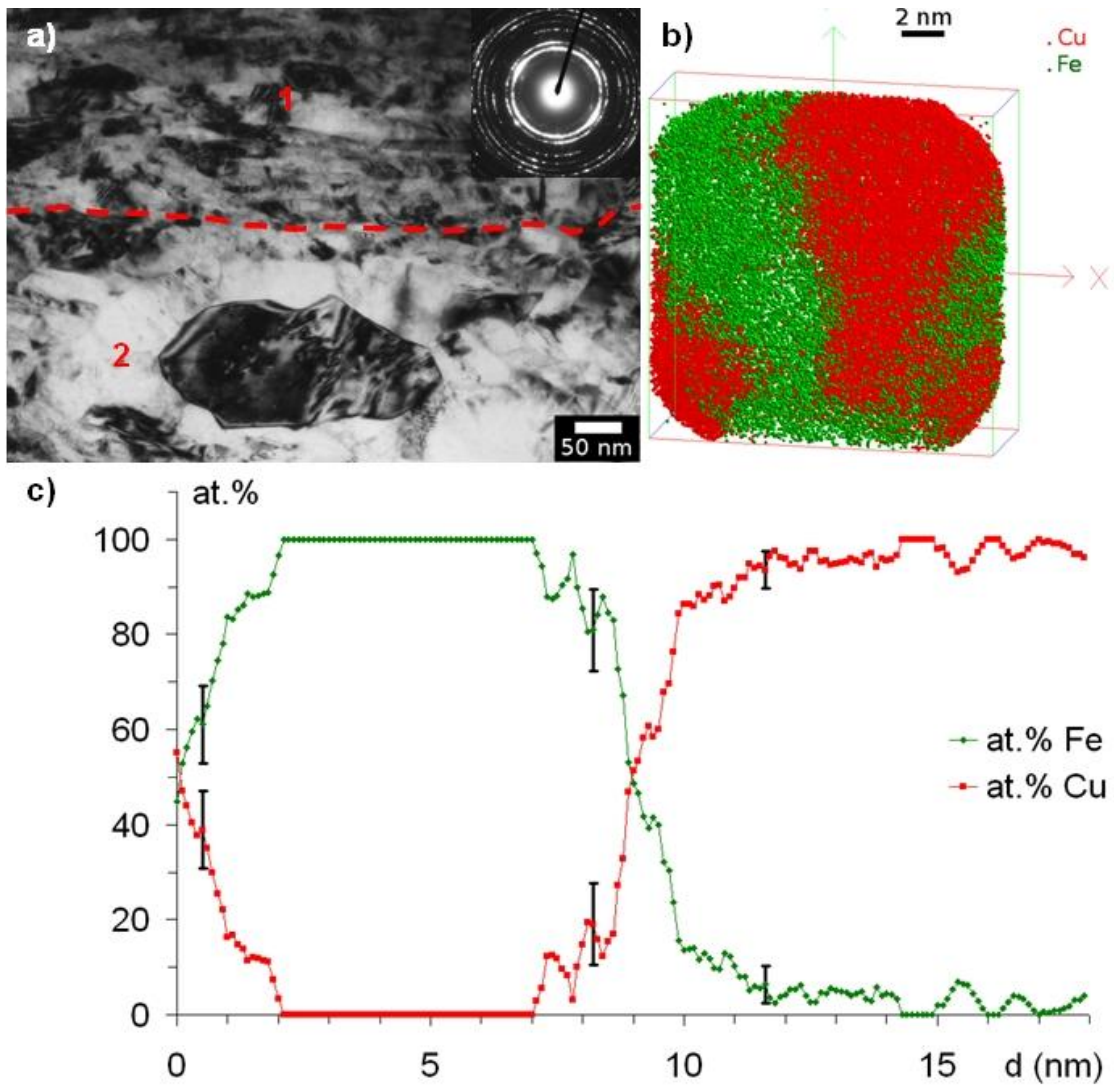

**Figure6.** (a) TEM bright field image and corresponding SAED pattern (inset) of the Cu-Fe nanocomposite processed 5 revolutions by HPT (at 293K with a strain rate of 0.2 s$^{-1}$). In the upper part of the image (labelled 1), the filamentary structure corresponds to a bunch of iron filaments where the bcc α-Fe is detected. In the lower part (labelled 2), the copper tube used for the first restacking is imaged. (b) Three-dimensional reconstruction of a volume analysed by APT in the filamentary region. Iron atoms are displayed in green and copper atoms in red. This image shows that the thickness of iron filaments is below 10 nm after 5 revolutions by HPT. (c) Composition profile computed across the iron filament displayed in the 3D volume (thickness of the sampling volume 0.5nm). A significant Fe gradient appears in the Cu matrix.

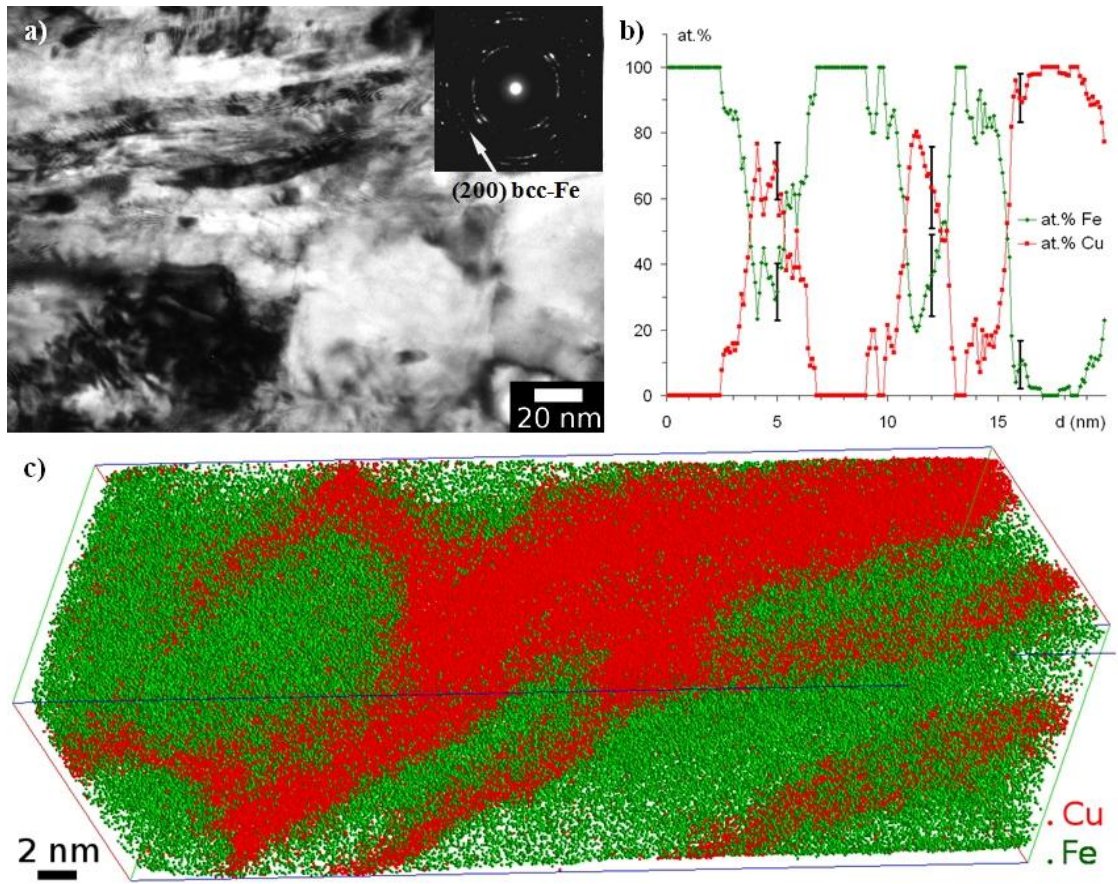

**Figure7.** (a) TEM bright field image and corresponding SAED pattern (inset) of the Cu-Fe nanocomposite processed 14 revolutions by HPT (at 293K with a strain rate of 0.2 s$^{-1}$). In some regions, a filamentary structure still exists with some bcc α-Fe. (c) Three-dimensional reconstruction of a volume analysed by APT. (b) Composition profile computed across iron filaments displayed in the 3D volume (thickness of the sampling volume 0.5nm). Most of iron filaments are not affected by the mechanical mixing, but a significant amount of Fe atoms (about 30at.%) is detected in solid solution in the Cu matrix.

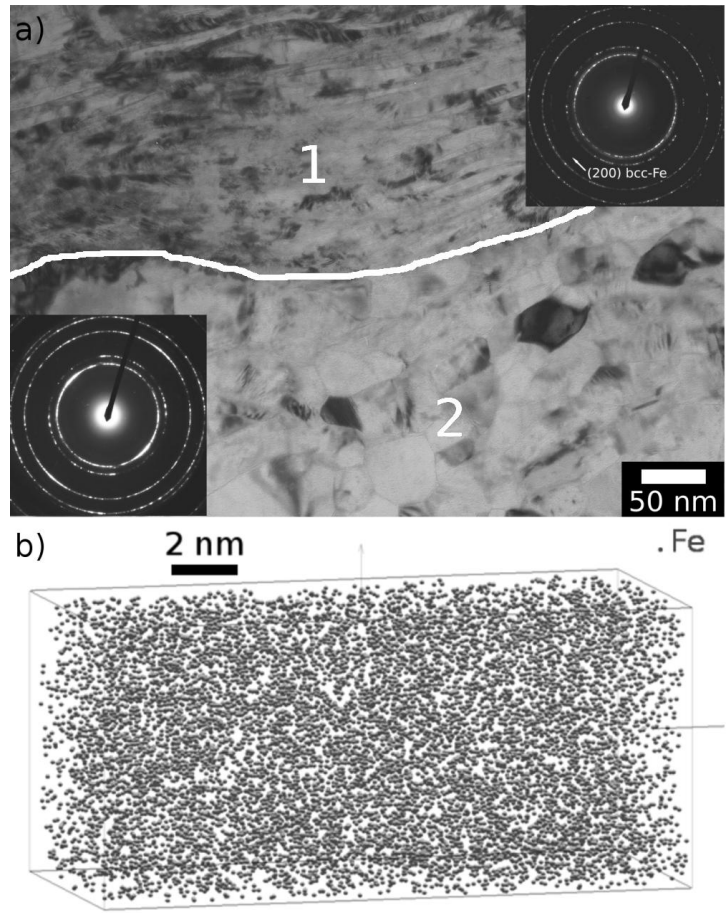

**Figure 8.** (a) TEM bright field image of the Cu-Fe nanocomposite processed 14 revolutions by HPT at 77K with a strain rate of 0.2 s$^{-1}$. In the region labelled 1, there is a filamentary structure and the bcc α-Fe is detected in the corresponding SAED pattern. In the region labelled 2, grains are equi-axed with an average size of about 50nm. Only the fcc Cu phase is detected on the corresponding SAED pattern. (b) Three-dimensional reconstruction of a volume analysed by APT in the Cu-Fe nanocomposite processed 25 revolutions by HPT at 77K with a strain rate of 0.2 s$^{-1}$. Only Fe atoms are displayed to show their random distribution. The Fe concentration in this analysed volume is 12.2 at.% (±0.1).